\documentclass[aps,twocolumn,pra,groupedaddress,superscriptaddress,nofootinbib,amsmath,amssymb,amsthm,notitle]{revtex4-1}

\usepackage{graphicx,amsmath,amssymb,txfonts,float}
\usepackage{subfigure,hyperref,bbm,times}
\usepackage[T1]{fontenc}

\usepackage{graphicx,amsmath,amssymb,txfonts,float}
\usepackage{subfigure,hyperref,bbm,times}
\usepackage[T1]{fontenc}
\usepackage{braket}
\usepackage{epsfig}
\usepackage{color}
\usepackage{graphicx}
\usepackage{dcolumn}
\usepackage{bm}

\providecommand{\openone}{\leavevmode\hbox{\small1\kern-3.8pt\normalsize1}}
\usepackage{soul}

\newcommand{\cwb}[1]{{\color{blue} \st{#1}}}

\begin{document}

\title{Indistinguishability-assisted two-qubit entanglement distillation}

\author{Farzam Nosrati}
\email{farzam.nosrati@unipa.it}
\affiliation{Dipartimento di Ingegneria, Universit\`{a} degli Studi di Palermo, Viale delle Scienze, 90128 Palermo, Italy}
\affiliation{INRS-EMT, 1650 Boulevard Lionel-Boulet, Varennes, Qu\'{e}bec J3X 1S2, Canada}

\author{Bruno Bellomo}
\affiliation{Universit\'{e} de Franche-Comt\'{e}, Institut UTINAM, CNRS UMR 6213, Observatoire des Sciences de
l'Univers THETA, 41 bis avenue de l'Observatoire, F-25010 Besan\c{c}on, France}

\author{Gabriele De Chiara}
\affiliation{Centre for Quantum Materials and Technology
, Queen's University Belfast, Belfast BT7 1NN, United Kingdom}
\author{Giuseppe Compagno}
\affiliation{Dipartimento di Fisica e Chimica - Emilio Segr\`e, Universit\`{a} degli Studi di Palermo, via Archirafi 36, 90123 Palermo, Italy}
\author{Roberto Morandotti}
\affiliation{INRS-EMT, 1650 Boulevard Lionel-Boulet, Varennes, Qu\'{e}bec J3X 1S2, Canada}
\author{Rosario Lo Franco}
\email{rosario.lofranco@unipa.it}
\affiliation{Dipartimento di Ingegneria, Universit\`{a} degli Studi di Palermo, Viale delle Scienze, 90128 Palermo, Italy}

\begin{abstract} 

Production of quantum states exhibiting a high degree of entanglement out of noisy conditions is one of the main goals of quantum information science. Here, we provide a conditional yet efficient entanglement distillation method which functions within the framework of spatially localized operations and classical communication. This method exploits indistinguishability effects due to the spatial overlap between two identical qubits in distinct sites and encompasses particle statistics imprint. We derive the general conditions for the maximum entanglement distillation out of mixed states. As applications, we give a thorough description of distilled entanglement and associated success probability starting from typical noisy states, such as thermal Gibbs states and Werner states. The influence of local temperatures and of noise parameter is discussed, respectively, in these two cases. The proposed scheme paves the way towards quantum repeaters in composite networks made of controllable identical quantum particles. 

\end{abstract}

\maketitle

\section{Introduction}

Indistinguishability plays an important role in understanding systems made of identical quantum entities. Usually, this genuinely quantum trait arises from the unaddressability of individual particles of the same kind when particles themselves become spatially overlapping \cite{peres2006quantum,tichy,franco2016quantum}. This property profoundly affects how these particles assemble to form composite quantum states of light and matter \cite{pauli2012general, griffiths2018introduction}. Beyond these fundamental aspects, it is known that tailoring spatial overlap of identical particles allows controllable entanglement generation schemes \cite{Bellomo2017n,franco2018indistinguishability,sun2020experimental,
barros2020entangling,lee2022entangling,piccolini2023asymptotically} and quantum information tasks, including teleportation \cite{franco2018indistinguishability,sun2020experimental}, quantum estimation \cite{castellini2019indistinguishability, Sun2022Activation}, and entanglement distribution between nodes of a quantum network \cite{castellini2019activating,Wang2022Remote}. Also, bosonic bunching, a symptom of indistinguishability, can enhance the conversion of information and work \cite{Bengtsson2018QuantumSzilard, watanabe2020quantum, holmes2020enhanced}. Recently, it has been reported that bosonic indistinguishability leads to an efficiency at maximum power larger than the classical Curzon-Ahlborn limit \cite{myers2022boosting}. 

Entanglement and quantum coherence are essential resources at the heart of quantum-enhanced technologies, including computation, secure communication \cite{Ekert1992Quantum}, and sensing \cite{giovannetti2011advances}. Nonetheless, noise and environmental interaction are inevitable, leading to decoherence and entanglement degradation \cite{breuer2002theory,aolita_2015}. Hence, quantum information and computation science faces the challenge of preserving quantum resources within a given noise model. Many protection schemes have been proposed to preserve such quantum resources \cite{maniscalco2008protecting,zanardi1997noiseless,Viola1998,
mortezapour2018protecting,xu_2010,bylicka_2014,man_2015,tan_2010,tong_2010,
piccolini2021opensys,Piccolini_2021_entropy,cuevas_2017,damodarakurup_2009,xu_2013,
lo_franco_2012_pra,facchi_2004,orieux_201,darrigo_2014_aop}. However, instead of maintaining entangled pairs, a distillation protocol (also referred to purification) converts mixed states to entangled states using local operations and classical operations (LOCC) performed by two local observers \cite{Bennett1996Mixed,Murao1998Multiparticle}. As a result, many distillation techniques have been introduced, aiming at increasing entanglement in quantum states utilizing LOCC, local filtering \cite{GISIN1996151Hidden}, distillation protocols \cite{Bennett1996Mixed}, noiseless linear amplification, and one-shot distillable entanglement \cite{Ecker2021Experimental}. Furthermore, entanglement distillation is a crucial feature of many proposals for quantum repeaters \cite{Briegel1998Quantum,Bratzik2013Quantum,Guha2015Rate}. 

In this work we provide a probabilistic entanglement distillation procedure based on spatial overlap between identical particles in the context of spatially localized operations and classical communication (sLOCC). We identify the essential conditions for a nonzero success rate of entanglement distillation. Due to the probabilistic nature of the distillation protocol, there is an interesting trade-off between the degree of distilled entanglement and the probability of success. Hence, we optimize the success probability of distillation at the expense of a lower entanglement degree. As relevant applications, we present a thorough examination of the requirements for entanglement distillation starting from thermal and Werner input states. Our procedure only uses inherent properties of identical particles as elementary building blocks of composite quantum networks. Therefore, it is particularly suitable in quantum communication scenarios, allowing for the conversion of seemingly useless noisy states into operational resource states for transferring information.

This paper is structured as follows. In Sec.~\ref{SecII} we describe the distillation protocol, examine some general properties of the final state, and present some quantities useful to study the trade-off between the amount of distilled entanglement and the success probability of the protocol. In Sec.~\ref{SecIII} and \ref{SecIV} we present our analysis of the protocol in the case of, respectively, thermal and Werner input states.
Finally, in Sec.~\ref{SecV} we draw our conclusions and present some remarks, while some details of our analysis can be found in the Appendixes.

\section{Distillation protocol}\label{SecII}
Before introducing the distillation protocol, let us briefly recall our primary framework for dealing with identical particles. We assume each particle to be endowed with an external and an internal degree of freedom, namely a spatial wave function and a pseudospin. In the no-label approach \cite{franco2016quantum,compagno2018dealing}, an elementary form of the two-particle state $\ket{\Psi}=\ket{\Phi_1,\Phi_2}$ is defined by single-particle state vectors $\ket{\Phi_i}=\ket{\psi_i\:\sigma_i}$ ($i=1,2$), where $\ket{\psi_i}$ is the spatial wave function and $\ket{\sigma}=\{\ket{\uparrow},\ket{\downarrow}\}$ is the pseudospin. 

\begin{figure}[t!] 
\centering
\includegraphics[width=0.48\textwidth]{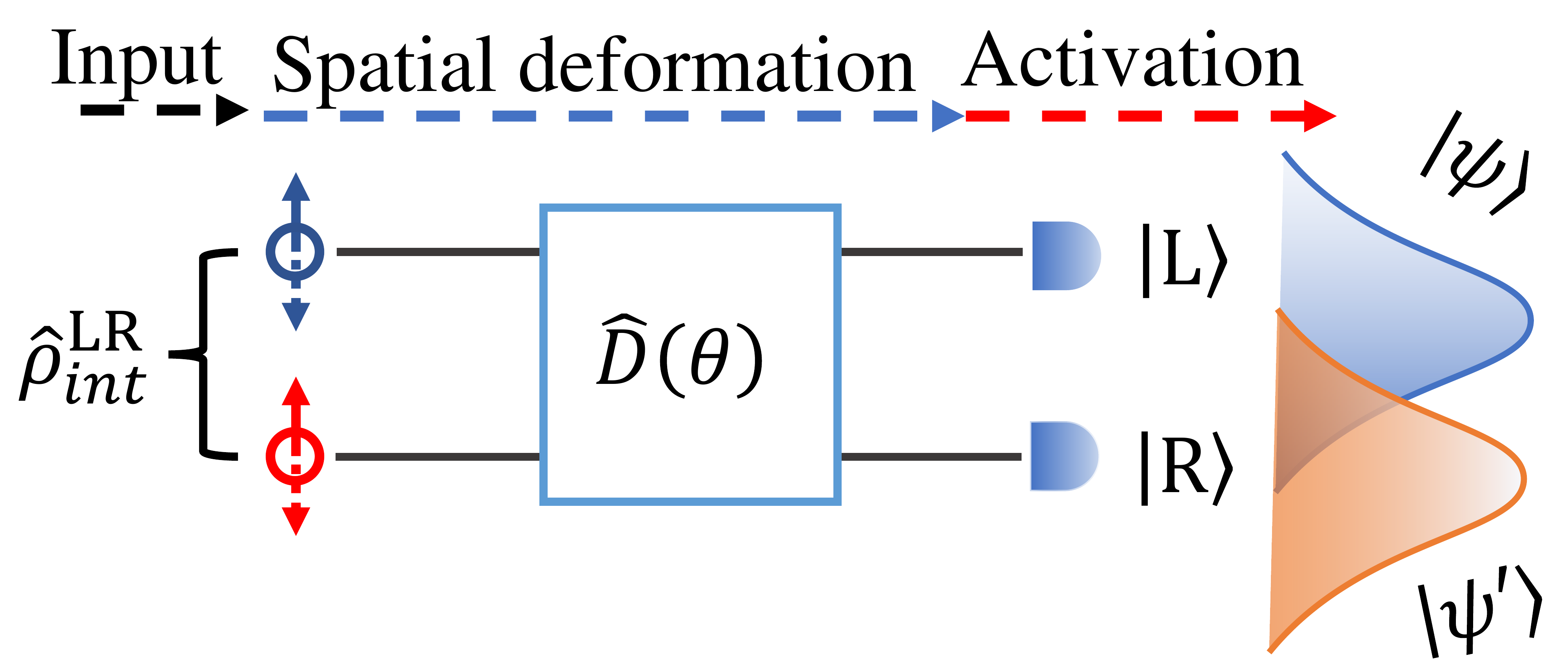}
\caption{\textbf{Scheme of indistinguishability-assisted entanglement distillation}. As an input, each (spatially distinguishable) qubit is independently prepared, so that $\hat{\rho}_\mathrm{init}^{\mathrm{LR}}=\hat{\rho}_{\mathrm{L}}\otimes\hat{\rho}_{\mathrm{R}}$. \textit{Step 1: spatial deformation}. An unitary spatial deformation $\hat{D}(\theta)$ is applied which allows the identical qubits to spatially overlap in the two distinct measurement regions. \textit{Step 2: activation}. The postselection measurement $\Pi^{(2)}_\mathrm{LR}$ is performed and entanglement is thus distilled.}
\label{Fig_1}
\end{figure}

As illustrated in Fig.~\ref{Fig_1}, the indistinguishability-assisted distillation protocol involves two steps: (i) an unitary transformation of the spatial wave functions, which we call \emph{spatial deformation} \cite{piccolini2022philtrans}; (ii) a postselection measurement typically involving classical communication between two observers (detectors) situated in separated regions L and R, which we name \emph{activation}. 

We assume that each qubit is initially in a diagonal state $\hat{\rho}_{\mathrm{X}}=\sum_{\sigma=\{\uparrow, \downarrow\}}\lambda_{\sigma}^{\mathrm{X}}\ket{\mathrm{X}\:\sigma}\bra{\mathrm{X}\:\sigma}$ in the basis $\{\ket{\mathrm{X}\:\sigma}\}$, with $\mathrm{X}=\{\mathrm{L},\mathrm{R}\}$ and $\sigma=\{\uparrow, \downarrow\}$, characterized by the probabilities of occurrence $\lambda_{\sigma}^{\mathrm{X}}$ ($\lambda_{\uparrow}^{\mathrm{X}}+\lambda_{\downarrow}^{\mathrm{X}}=1$). Since the qubits are initially distinguishable in two distinct sites, they are individually addressable by their spatial modes. Thus, the initial two-particle state is separable and factorizable in the two spatial regions, that is $\hat{\rho}_\mathrm{init}^{\mathrm{LR}}=\hat{\rho}_{\mathrm{L}}\otimes\hat{\rho}_{\mathrm{R}}$. This state can be written as 
\begin{equation}\label{eq: rho_seprable}
\hat{\rho}_\mathrm{init}^{\mathrm{LR}}=\sum_{\sigma,\tau=\{\uparrow, \downarrow\}}\lambda_{\sigma,\tau}^{\mathrm{LR}}\ket{\mathrm{L}\:\sigma,\mathrm{R}\:\tau}\bra{\mathrm{L}\:\sigma,\mathrm{R}\:\tau},
\end{equation}
where the probability coefficients are $\lambda_{\sigma,\tau}^{\mathrm{LR}}=\lambda_{\sigma}^{\mathrm{L}}\lambda_{\tau}^{\mathrm{R}}$. We show below how two observers located in two spatial distinct regions can accomplish entanglement distillation (purification), by performing spatially local particle counting on the shared pair of identical qubits, and coordinating their actions through classical messages if needed. 

\textbf{Step I: spatial deformation.} The spatial deformation involves a transformation of the spatial wave functions which distributes them onto the two operational spatial sites L and R. In this way, the identical qubits can spatially overlap and are not localized anymore in their own respective regions. 
In an operational context, we consider the case of an input state experiencing a spatial deformation transformation, in which only the shapes of single-particle wave functions are altered, leaving the internal pseudospin degrees of freedom unchanged. To do so, we perturb the system by means of a spatial deformation Hamiltonian, defined as (hereafter we take $\hbar=1$)
\begin{equation}
    \hat{H}_D=\frac{\Omega_R}{2}\sum_{\sigma=\{\uparrow, \downarrow\}}\left(e^{\mathbf{i}\phi}\hat{L}_\sigma^{\dagger}\hat{R}_\sigma+e^{-\mathbf{i}\phi}\hat{L}_{\sigma}\hat{R}^{\dagger}_\sigma\right),
\end{equation}
where $\Omega_R$ is the coupling strength and $\phi$ is an arbitrary phase. Here, $\hat{X}_\sigma^{\dagger}$ and $\hat{X}_\sigma$ ($\hat{X}=\{\hat{L},\hat{R}\}$) are, respectively, the creation and the annihilation operators in the site $\mathrm{X}=\{\mathrm{L},\mathrm{R}\}$, with the action of $\hat{X}_\sigma^{\dagger}$ on the vacuum state $\ket{0}$ given by  $X_\sigma^{\dagger}\ket{0}=\ket{\mathrm{X}\:\sigma}$. These creation and annihilation operators obey standard anticommutation rules,  $\{\hat{X}_\sigma^{\dagger},\hat{X}_\sigma'^{\dagger}\}=\delta_{\sigma,\sigma'}$ for fermions, and commutation rules, $[\hat{X}_\sigma,\hat{X}_\sigma'^{\dagger}]=\delta_{\sigma,\sigma'}$  for bosons \cite{cohen2006quantum}. These rules make the two-particle state fulfil the standard phase change associated to a particle exchange, $\ket{\mathrm{L}\:\sigma,\mathrm{R}\:\tau}=\eta\ket{\mathrm{R}\:\tau,\mathrm{L}\:\sigma}$ with $\eta=1(-1)$ for bosons (fermions), and, therefore, the Pauli exclusion principle for fermions \cite{pauli2012general}. The time-evolution operator associated to the spatial deformation is  $\hat{D}(\theta)=\exp{(-\mathbf{i}t\hat{H}_D )}$, with $\theta=\Omega_Rt$, so that the initial state evolves as $\hat{\rho}_D=\hat{D}(\theta)\hat{\rho}_{int}^{\mathrm{LR}}\hat{D}^\dagger(\theta)$. As a consequence, the deformed state has the form
\begin{equation} \label{eq: deformed_state}
    \hat{\rho}_{D}=\sum_{\sigma,\tau=\{\uparrow, \downarrow\}}\lambda_{\sigma,\tau}^{\mathrm{LR}}\ket{\psi\:\sigma,\psi'\:\tau}\bra{\psi\:\sigma,\psi'\:\tau},
\end{equation}
where $\ket{\psi}$ and $\ket{\psi'}$ are the deformed spatial wave functions defined as [see \ref{appendix1} for the derivation of Eq.~$($\ref{eq: deformed_state}$)$]
\begin{eqnarray}\label{spatialdefor}
\ket{\psi}&=&\cos{\left(\frac{\theta}{2}\right)}\ket{\mathrm{L}}+\mathbf{i}e^{\mathbf{i}\phi}\sin{\left(\frac{\theta}{2}\right)}\ket{\mathrm{R}}, \nonumber\\
\ket{\psi'}&=&\mathbf{i}e^{-\mathbf{i}\phi}\sin{\left(\frac{\theta}{2}\right)}\ket{\mathrm{L}}+\cos{\left(\frac{\theta}{2}\right)}\ket{\mathrm{R}}.
\end{eqnarray}
After this deformation, one may find each of the two particles in both regions, allowing them to spatial overlap. With spatial overlap we mean that the particles can be found in the same region of space with nonzero probability. It is worth mentioning that fermions with the same pseudospin are always forced to be in different sites due to the Pauli exclusion principle: the only accessible state in this case becomes, up to a global phase, $\ket{\mathrm{L}\:\sigma,\mathrm{R}\:\sigma}$.

\textbf{Step II: activation.} In the last step, the two observers in regions $\{\mathrm{L},\mathrm{R}\}$ distillate or purify exploitable quantum states by selecting only one particle per spatial region, via particle projective measurements $\hat{\Pi}_{\mathrm{LR}}^{(2)}=\sum_{\sigma,\tau=\uparrow,\downarrow}\ket{\mathrm{L}\:\sigma,\mathrm{R}\:\tau}\bra{\mathrm{L}\:\sigma,\mathrm{R}\:\tau}$ \cite{franco2018indistinguishability}. The procedure is probabilistic since it discards the cases when both particles are detected in the same site. The action of projective measurements on the deformed state of Eq.~$($\ref{eq: deformed_state}$)$, that is $\hat{\rho}^\mathrm{LR}_\mathrm{fin}=\hat{\Pi}_\mathrm{LR}^{(2)}\hat{\rho}_\mathrm{D}\hat{\Pi}_\mathrm{LR}^{(2)}/\mathrm{Tr}(\hat{\Pi}_\mathrm{LR}^{(2)}\hat{\rho}_\mathrm{D})$, gives
\begin{equation}
\label{eq: rho_LR}
\hat{\rho}^\mathrm{LR}_\mathrm{fin}=\sum_{\sigma_1,\sigma_2,\tau_1,\tau_2=\{\uparrow, \downarrow\}}\Lambda_{\sigma_1,\tau_1}^{\sigma_2,\tau_2}\ket{\mathrm{L}\:\sigma_1,\mathrm{R}\:\tau_1}\bra{\mathrm{L}\:\sigma_2,\mathrm{R}\:\tau_2},
\end{equation}
which is written in the computational basis $\mathcal{B}^\mathrm{LR}=\{\ket{\mathrm{L}\:\uparrow,\mathrm{R}\:\uparrow}, \ket{\mathrm{L}\:\uparrow,\mathrm{R}\:\downarrow}, \ket{\mathrm{L}\:\downarrow,\mathrm{R}\:\uparrow},$ $ \ket{\mathrm{L}\:\downarrow,\mathrm{R}\:\downarrow}\}$. The coefficients $\Lambda_{\sigma_1,\tau_1}^{\sigma_2,\tau_2}$ of Eq.~$($\ref{eq: rho_LR}$)$ depend on  initial probabilities, spatial deformation, and both fermionic and bosonic statistics (see \ref{appendix2} for a full description of the coefficients $\Lambda_{\sigma_1,\tau_1}^{\sigma_2,\tau_2}$). There is a success probability $P_\mathrm{LR}=\mathrm{Tr}(\hat{\Pi}_\mathrm{LR}^{(2)}\hat{\rho}_\mathrm{D})$
associated to the case when only one particle is found in a given region. The entanglement distillation protocol is thus conditional and encompasses the possibility of failure. We remark that classical communication between the two observers is typically involved in the protocol to check if each of them actually finds one particle per region. We also point out that the post-selected state is independent of the phase $\phi$.

In general, indistinguishability depends on both the quantum state and the measurement performed on the system \cite{nosrati2020robust}. Since we are interested in the condition of spatial overlap between two identical qubits in two distinct regions, our goal is to determine whether the qubits are spatially distinguishable. To this aim, we can use the joint sLOCC measurement $\hat{\Pi}_\mathrm{LR}^{(2)}$, to evaluate the indistinguishability due to the spatial overlap of the particles within the state by means of the success probability  $P_\mathrm{LR}=\mathrm{Tr}(\hat{\Pi}_\mathrm{LR}^{(2)}\hat{\rho}_\mathrm{D})$. If  $P_\mathrm{LR}=1$, indicating the presence of a single particle in each region with certainty, then the particles are considered spatially distinguishable with no spatial overlap. Conversely, if $P_\mathrm{LR}< 1$, the particles overlap in the two regions and are considered to be spatially indistinguishable. Once fixed the sLOCC measurement, the success probability depends in our setup on the initial state and on the parameter $\theta$ governing the spatial overlap.

Before considering specific input states from two different physical scenarios of interest, we examine some general properties of the final state of Eq.~$($\ref{eq: rho_LR}$)$. We first observe that for equally deformed spatial wave functions in the case of bosons, i.e., for $\theta=\pi/2$ in Eq.~$($\ref{spatialdefor}$)$ and $\eta=1$, the output state reduces to a pure maximally entangled state in the singlet state form, $\ket{\Psi^\mathrm{LR}_-}=\left(\ket{\mathrm{L}\:\uparrow,\mathrm{R}\:\downarrow}-\ket{\mathrm{L}\:\downarrow,\mathrm{R}\:\uparrow}\right)/\sqrt{2}$, with a success probability given by
\begin{equation}\label{eq: P_LR max}
P_\mathrm{LR}=\frac{\lambda_{\uparrow,\downarrow}^{\mathrm{LR}}+\lambda_{\downarrow,\uparrow}^{\mathrm{LR}}}{2}.
\end{equation}
The above expression implies that the occurrence of maximum entanglement distillation is conditional to the presence in the initial state of at least one term with opposite pseudospins. In fact, the success probability of Eq.~$($\ref{eq: P_LR max}$)$ tells us that not every mixed state is distillable to a pure entangled singlet state, in line with previous observations \cite{Horodecki1998mixed}. Furthermore, the upper bound for the success probability of maximum entanglement distillation is $P_\mathrm{LR}=1/2$ (when $\lambda_{\uparrow,\downarrow}^{\mathrm{LR}}+\lambda_{\downarrow,\uparrow}^{\mathrm{LR}}=1$), independently of the purity of the input state. Interestingly, the same result does not apply to fermions, for which the degree of distilled entanglement depends on the input state. The general final state in the case of  fermions is described by Eq.~$($\ref{eq: B1}$)$. Independently from the value of $\theta$, the output state depends on the values of $\lambda_{\uparrow,\uparrow}^{\mathrm{LR}}$ and $\lambda_{\downarrow,\downarrow}^{\mathrm{LR}}$, which are connected to the initial state. Consequently, for example, in the case of an input state with nonzero values for at least one of these two coefficients, maximum entanglement distillation becomes unattainable.

More in general, there is a trade-off between the amount of distillable entanglement and the likelihood of success of the procedure. This aspect could be exploited in a practical scenario where the source is probabilistic and increasing the distillation success probability is demanding. In fact, one can optimize the success probability by distilling non-maximal entanglement: sacrificing perfect entanglement in favor of success probability can be strategical in many realistic scenarios. 

We remark that the profound quantum property activating this entanglement distillation is the indistinguishability of the identical qubits stemming from their established spatial overlap in two distinct sites after spatial deformation of the wave functions.

Many ways can be employed to characterize the amount of distilled entanglement. For example, it can be quantified by the sLOCC-based concurrence, $C(\hat{\rho}^\mathrm{LR})=\max \{0,\sqrt{\lambda_4}-\sqrt{\lambda_3}-\sqrt{\lambda_2}-\sqrt{\lambda_1}\}$, used for identical particles \cite{nosrati2020robust}, where $\lambda_i$ are the eigenvalues in decreasing order of the operator $\hat{R}=\hat{\rho}^\mathrm{LR}\Tilde{\hat{\rho}}^\mathrm{LR}$, being $\Tilde{\hat{\rho}}^\mathrm{LR}=\hat{\sigma}^\mathrm{L}_{y}\otimes\hat{\sigma}_{y}^\mathrm{R}(\hat{\rho}^{\mathrm{LR}})^*\hat{\sigma}_{y}^\mathrm{L}\otimes\hat{\sigma}_{y}^\mathrm{R}$, with localized $y$-Pauli matrices $\hat{\sigma}_{y}^\mathrm{X}=\ket{\mathrm{X}}\bra{\mathrm{X}}\otimes\hat{\sigma}_{y}$ ($\mathrm{X}=\mathrm{L}, \mathrm{R}$). Notice that this sLOCC-based concurrence is directly connected to the usual concurrence defined for distinguishable qubits \cite{concurrence}.

Interestingly, one may want to ensure that the procedure will distillate enough entanglement to violate a Bell inequality. According to the Horodecki criterion for the CHSH-Bell inequality violation of a two-qubit density matrix \cite{HORODECKI1995340}, the expression of the maximum of the Bell function for  the $X$-shape density
matrix $\hat{\rho}^\mathrm{LR}_\mathrm{fin}$ can be written as $B_\mathrm{LR}=2\sqrt{\mathcal{P}^2+\mathcal{Q}^2}$ with $\mathcal{P}=\Lambda_{\uparrow,\uparrow}^{\uparrow,\uparrow}+\Lambda_{\downarrow,\downarrow}^{\downarrow,\downarrow}-\Lambda_{\uparrow,\downarrow}^{\uparrow,\downarrow}-\Lambda_{\downarrow,\uparrow}^{\downarrow,\uparrow}$ and $\mathcal{Q}=2\Lambda_{\uparrow,\downarrow}^{\downarrow,\uparrow}$, where we used the fact that here $\Lambda_{\uparrow,\downarrow}^{\downarrow,\uparrow}$ is real and $\Lambda_{\uparrow,\uparrow}^{\downarrow,\downarrow}=\Lambda_{\downarrow,\downarrow}^{\uparrow,\uparrow}=0$. In particular, the condition $B_\mathrm{LR}>2$ is sufficient to perform quantum teleportation. 

On the other hand, one may find a mixed bipartite state that does not violate any Bell inequality but is still useful for quantum teleportation \cite{popescu1994Bell}. A possible way to also include these cases is to characterize the amount of entanglement using the state's distance to an ideal one, such as the singlet state $\ket{\Psi^\mathrm{LR}_-}$, in terms of the fidelity $F_\mathrm{LR}(\hat{\rho}^\mathrm{LR}_\mathrm{fin})=\bra{\Psi^\mathrm{LR}_-}\hat{\rho}^\mathrm{LR}_\mathrm{fin}\ket{\Psi^\mathrm{LR}_-}$. For instance, for bosons when $\theta=\pi/2$ the fidelity is maximum, i.e., $F_\mathrm{LR}=1$, and the success probability is given by Eq.~$($\ref{eq: P_LR max}$)$. However, a fidelity $F_\mathrm{LR}>2/3$ is enough for achieving quantum teleportation \cite{hu2013relations}. In such a case, one may prefer a higher probability of success at the expense of a lower fidelity. 

As a result of the above considerations, the entanglement distillation analysis is a constrained optimization problem in which we seek to maximize the success probability with respect to $\theta$, $P_\mathrm{LR}^{\max}=\stackrel[\theta]{}{\max}\left[{\mathrm{Tr}(\hat{\Pi}_\mathrm{LR}^{(2)}\hat{\rho}_\mathrm{D})}\right]$, with the constraint $B_\mathrm{LR}(\hat{\rho}^\mathrm{LR}_\mathrm{fin})>2$ (for the Bell inequality threshold) or $F_\mathrm{LR}(\hat{\rho}^\mathrm{LR}_\mathrm{fin})>2/3$ (for the fidelity threshold). 

In the following sections, we shall analyze the performance of the indistinguishability-assisted entanglement distillation in the case of two well-known input states: a thermal state and a Werner state. 

\section{Application I: Thermal state}\label{SecIII}

As a first example, we consider the distillation of thermal Gibbs states of two spatially separated identical qubits. From a resource theory point of view, a thermal state is a free state that one can generate and use at no cost. Therefore, one can interpret the distillation process as a quantum thermal machine that produces entanglement out of a thermal state. 

To define the initial thermal state, we consider the bare system Hamiltonian $\hat{H}_0^{\mathrm{X}}=\frac{\omega}{2}\hat{\sigma}_{z}^\mathrm{X}$ ($X=\mathrm{L},\mathrm{R}$), where $\hat{\sigma}_{z}^\mathrm{X}=\hat{\sigma}_{z}\otimes\ket{X}\bra{X}$ is the localized $z$-Pauli matrix, and $\omega=\omega_\uparrow-\omega_\downarrow$ is the frequency spacing between the energy levels associated to the values of the pseudospin. Each qubit is initially coupled with its own local bosonic bath $\mathcal{B}_\mathrm{X}$  at inverse temperature $\beta_X=1/T_\mathrm{X}$ (hereafter we take the Boltzmann's constant $k_B=1$). Thus, each particle thermal state is in the Gibbs form $\hat{\rho}_\mathrm{X}=e^{-\beta_X\hat{H}_0^{\mathrm{X}}}/\mathcal{Z}(\beta_\mathrm{X})$, where $\mathcal{Z}(\beta_\mathrm{X})={\rm Tr}\left(e^{-\beta_X\hat{H}_0^{\mathrm{X}}}\right)$ is the partition function. Therefore, the initial two-particle state is factorizable in terms of single-particle states, $\hat{\rho}^{\mathrm{LR}}_\mathrm{init}=\hat{\rho}_{\mathrm{L}}\otimes\hat{\rho}_{\mathrm{R}}$. 

\begin{figure}[t!] 
	\centering
	\includegraphics[width=0.48\textwidth]{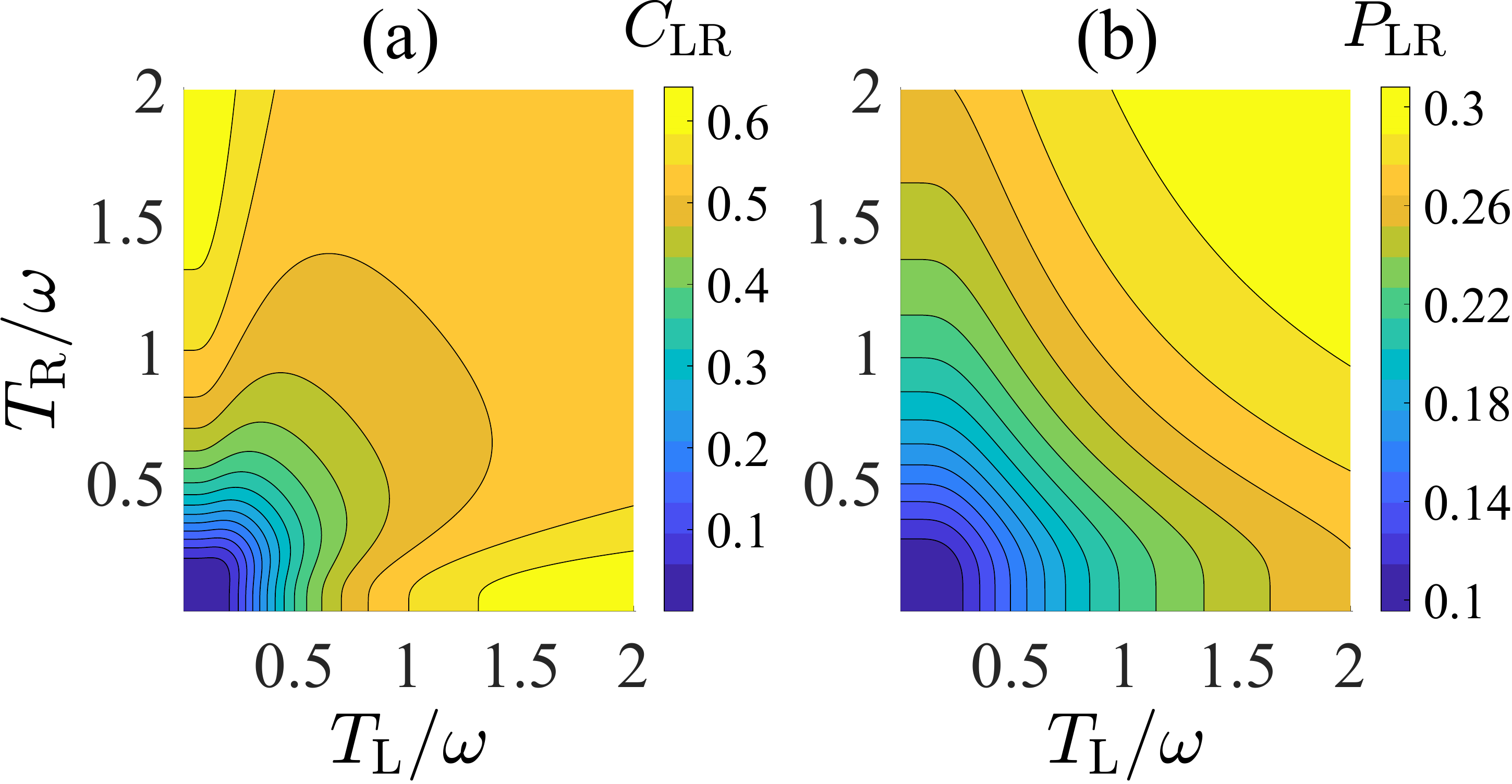}
	\caption{\textbf{Distilled entanglement and success probability for thermal input states  with baths at different temperatures.}  (a) Counter plot of the concurrence $C_\mathrm{LR}$ and (b) success probability $P_\mathrm{LR}$ as functions of the temperatures $T_\mathrm{L}/\omega$ and $T_\mathrm{R}/\omega$  for $\theta=2\pi/5$, in the case of bosons.}
	\label{Fig:contour_plot}
\end{figure}

For the sake of simplicity, in the next analytical considerations we limit ourselves to the case of equal temperature baths, that is $\beta=\beta_{\mathrm{L}}=\beta_{\mathrm{R}}$ (the case of different temperatures will be explicitly considered via a computational analysis shown in Fig.~\ref{Fig:contour_plot}). After spatial deformation and projection, the activated output state of Eq.~$($\ref{eq: rho_LR}$)$ becomes
\begin{equation}\label{eq: G thermal}
    \hat{G}_\mathrm{fin}^\mathrm{LR}=\sum_{i=\{0, 1,\pm\}}\mathcal{Z}_{\mathrm{LR}}^{-1}\Lambda^\mathrm{LR}_{i}\ket{\Psi_i^{\mathrm{LR}}}\bra{\Psi_i^{\mathrm{LR}}},
\end{equation}
which is written in the singlet-triplet basis $\mathcal{B}^\mathrm{LR}_{\textrm{st}}=\{\ket{\Psi_0^{\mathrm{LR}}},\ket{\Psi_1^{\mathrm{LR}}},\ket{\Psi_\pm^{\mathrm{LR}}}\}$, where $\ket{\Psi_0^{\mathrm{LR}}}:=\ket{\mathrm{L}\:\downarrow,\mathrm{R}\:\downarrow}$, $ \ket{\Psi_1^{\mathrm{LR}}}:=\ket{\mathrm{L}\:\uparrow,\mathrm{R}\:\uparrow}$, $\ket{\Psi_\pm^{\mathrm{LR}}}:=\left(\ket{\mathrm{L}\:\downarrow,\mathrm{R}\:\uparrow}\pm\ket{\mathrm{L}\:\uparrow,\mathrm{R}\:\downarrow}\right)/\sqrt{2}$, and $\mathcal{Z}_{\mathrm{LR}}=\sum_{i=\{0,1,\pm\}}\Lambda^\mathrm{LR}_{i}$ [see \ref{appendix2} for some details concerning the derivation of Eq.~$($\ref{eq: G thermal}$)$]. The thermal coefficients for fermions are $\Lambda^\mathrm{LR}_{0}=e^{\beta\omega}$, $\Lambda^\mathrm{LR}_{1}=e^{-\beta\omega}$, $\Lambda^\mathrm{LR}_{+}=1$, and $\Lambda^\mathrm{LR}_{-}=\cos^2{(\theta)}$. The thermal coefficients for bosons are  $\Lambda^\mathrm{LR}_{0}=e^{\beta\omega}\cos^2{(\theta)}$, $\Lambda^\mathrm{LR}_{+}=\cos^2{(\theta)}$, and $\Lambda^\mathrm{LR}_{-}=1$. 

\begin{figure}[t!] 
	\centering
	\includegraphics[width=0.48\textwidth]{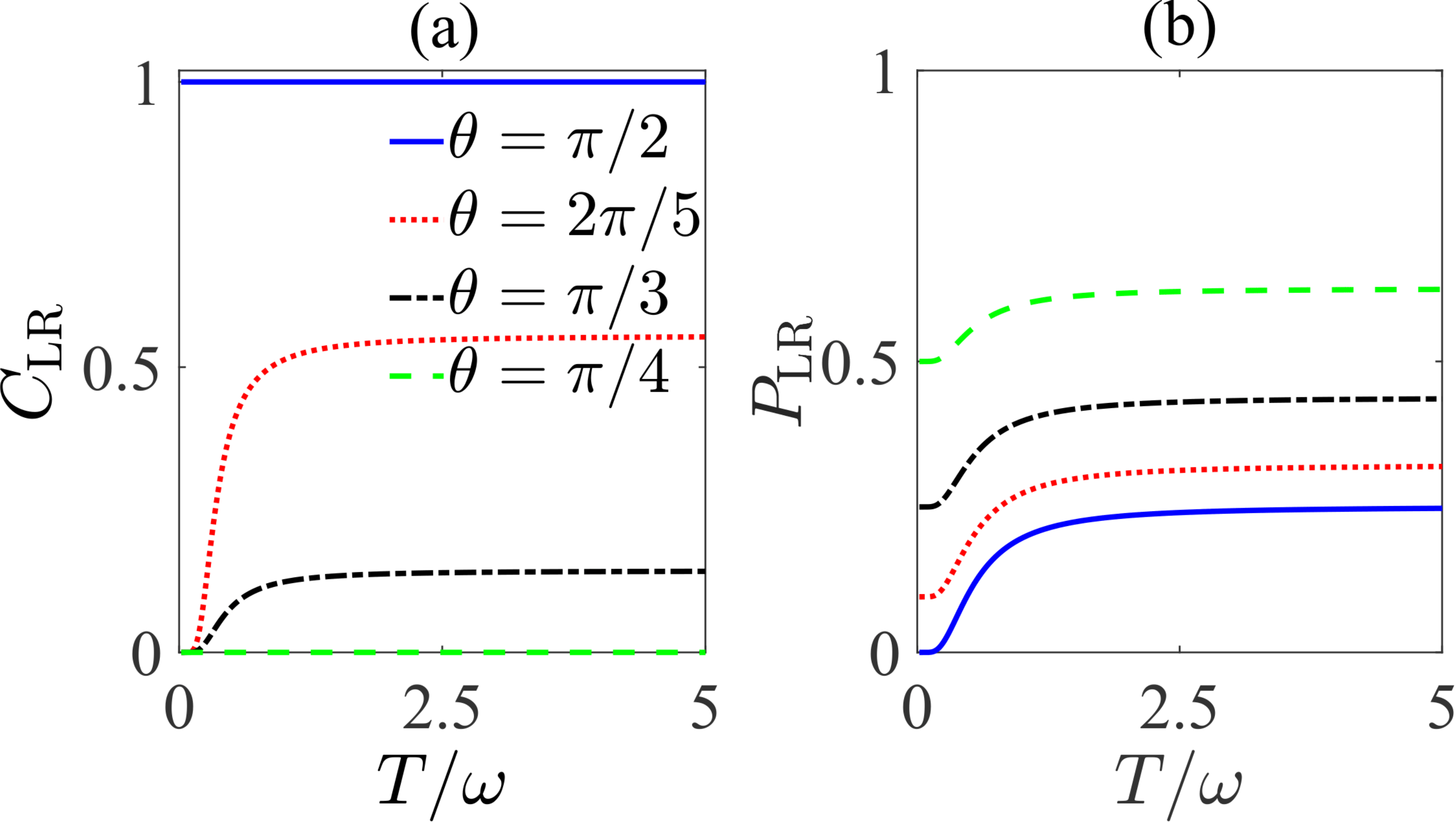}
	\caption{\textbf{Distilled entanglement and success probability for thermal input states  with baths at the same temperature.} (a) Concurrence $C_\mathrm{LR}=C(\hat{G}_\mathrm{fin}^\mathrm{LR})$ and (b) success probability $P_\mathrm{LR}$ as functions of the baths' temperature $T/\omega$ for different values of $\theta$, in the case of bosons.}
	\label{Fig: Thermal entanglement}
\end{figure}

We now provide some analytical results for both bosons and fermions. To quantify the amount of distilled entanglement out of the thermal state we use the concurrence $C$. For bosons ($\eta=1$), one gets $C(\hat{G}_\mathrm{fin}^{\mathrm{LR}})=\max \left\{0, \mathcal{C}\right\}$ with
\begin{equation}\label{eq: C_LR}
    \mathcal{C}=\frac{2-6 \cos^2{(\theta)}}{4 \cos^2{(\theta )} \cosh{(\beta\omega)}+\cos{(2\theta )}+3},
\end{equation}
while the success probability is equal to
\begin{equation}  \label{eq: P_sLOCC}
    P_\mathrm{LR}=1-\left[\frac{1+2\cosh{(\beta\omega)}}{2+2\cosh{(\beta\omega)}}\right]\sin^2{(\theta)}.
\end{equation}
Both $\mathcal{C}$ and $P_\mathrm{LR}$ range in $[0,1]$ and depend on the bath temperature and the spatial deformation parameter $\theta$. Differently, it is possible to show that for fermions ($\eta=-1$), the entanglement amount is always zero, although the state is non-diagonal in the computational basis $\mathcal{B}^\mathrm{LR}$. 

For spatially distinguishable qubits (i.e., for $\theta=0$), the output is the separable state expressed in Eq.~$($\ref{eq: rho_seprable}$)$ for both bosons and fermions. For an equally weighted spatial deformation of the wave functions ($\theta=\pi/2$) in the case of bosons, the state of Eq.~$($\ref{eq: G thermal}$)$ reduces to the maximally entangled singlet state $\ket{\Psi_-^{\mathrm{LR}}}$, as previously observed, with a success probability $P_\mathrm{LR}=1/[2+2\cosh{(\beta \omega)}]$, that ranges between $0$ and $1/4$. For $\theta=\pi/2$ in the case of fermions, the singlet state is ruled out and the state of Eq.~$($\ref{eq: G thermal}$)$ becomes $\hat{G}_f^\mathrm{LR}=\sum_{i=\{0,1,+\}}\mathcal{Z}_{\mathrm{LR}}^{-1}\Lambda_{i}^{\mathrm{LR}}\ket{\Psi_i^{\mathrm{LR}}}\bra{\Psi_i^{\mathrm{LR}}}$, with $\Lambda_{0}^{\mathrm{LR}}=e^{\beta\omega}$, $\Lambda_{1}^{\mathrm{LR}}=e^{-\beta\omega}$, and $\Lambda_{+}^{\mathrm{LR}}=1$. 

Using Eqs.~\ref{eq: C_LR} and \ref{eq: P_sLOCC}, valid for bosons, we display in Fig.~\ref{Fig: Thermal entanglement} $C_\mathrm{LR}=C(\hat{G}_\mathrm{fin}^\mathrm{LR})$ [panel (a)] and $P_\mathrm{LR}$ [panel (b)] as functions of the bath temperature for some values of $\theta$, i.e., for different settings of the spatial deformation of the wave functions. As commented above, for $\theta=\pi/2$ entanglement is always maximum irrespective of the bath temperature. For the other cases, contrarily to common expectations, the amount of entanglement increases with the temperature until it reaches a stationary value. This behavior can be explained with the fact that increasing the bath temperature leads to populating two-qubit states which exhibit quantum correlations arising from indistinguishability effects due to the spatial overlap. Instead, decreasing the temperature increases the likelihood of finding qubits in their ground state $\ket{\mathrm{L}\:\downarrow,\mathrm{R}\:\downarrow}$, which is a pure separable state from which entanglement cannot be distilled. The plot of the corresponding success probabilities shows that the process is ineffective for very low temperatures and that  curves with smaller degrees of distilled entanglement are linked to curves with larger values of $P_\mathrm{LR}$. 

We have seen that the amount of the distilled entanglement depends on the baths' temperature $T=T_\mathrm{L}=T_\mathrm{R}$. However, it may be useful to understand how the indistinguishability-assisted entanglement distillation procedure works in the case when the two baths have different temperatures. Fig.~\ref{Fig:contour_plot} shows a contour plot of the distilled entanglement, quantified by the concurrence $C_\mathrm{LR}$, and its corresponding  success probability $P_\mathrm{LR}$ in terms of two temperatures $T_\mathrm{L}$ and $T_\mathrm{R}$ in the case of a partial spatial deformation corresponding to $\theta=2\pi/5$. As an interesting aspect, one can notice that by increasing the difference between the two temperatures it is possible to augment the degree of the distilled entanglement.

\begin{figure}[t!] 
	\centering
	\includegraphics[width=0.48\textwidth]{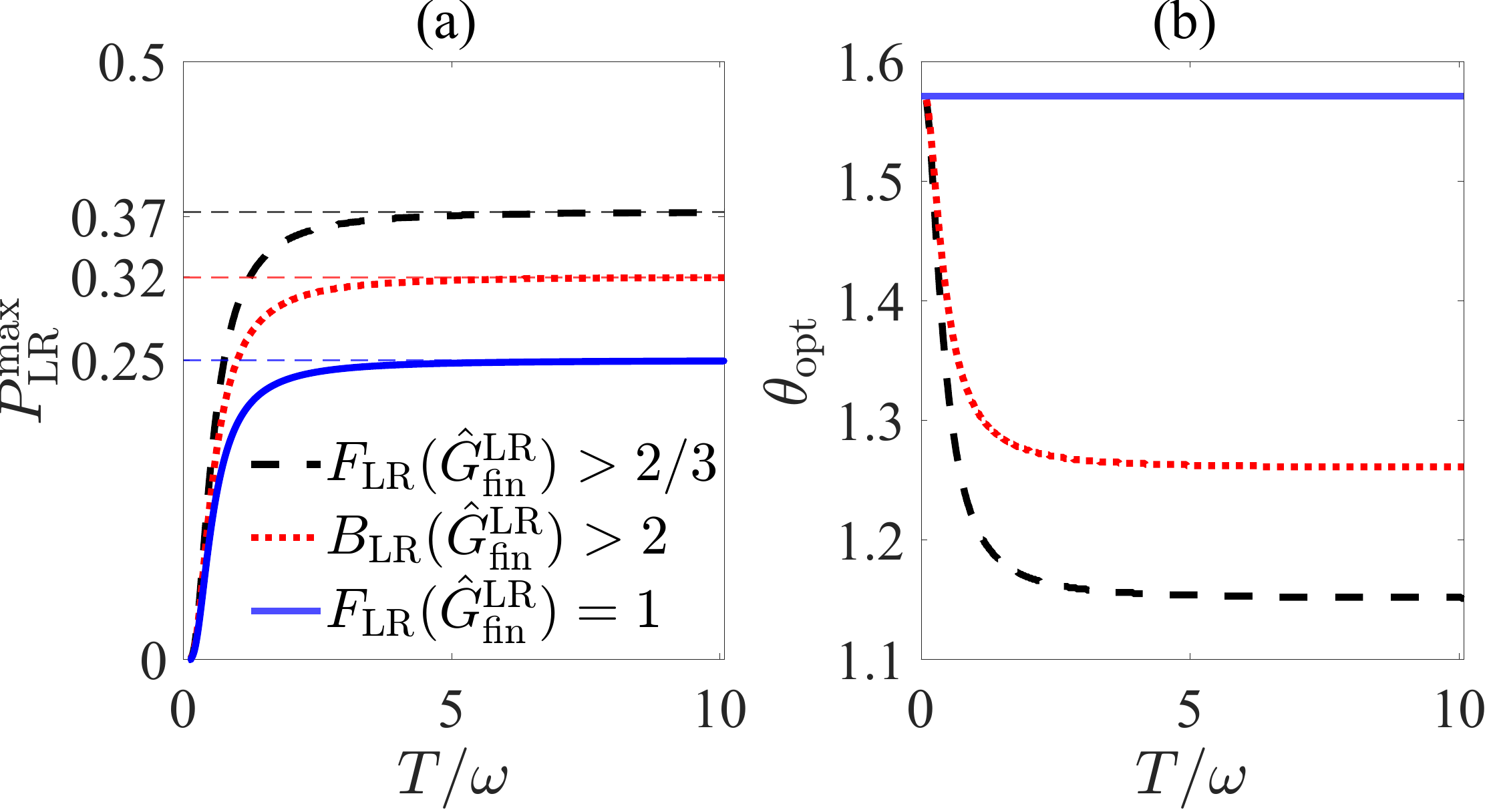}
	\caption{\textbf{Maximum success probability of entanglement distillation with thermal input states in baths at the same temperature.} (a) Maximum  success probability $P_\mathrm{LR}^{\max}$ as a function of $T/\omega$ for bosons subject to the constraints $F_\mathrm{LR}(\hat{G}^\mathrm{LR})>2/3$ (blue solid line)\cwb{)} and $B_\mathrm{LR}(\hat{G}^\mathrm{LR})>2$ (red dotted line). The black dashed line shows the success probability in the case of perfect entanglement distillation. (b) The corresponding optimized spatial deformation parameters $\theta_\mathrm{opt}$ as a function of $T/\omega$.}
	\label{Fig: maximum sLOCC}
\end{figure}

To conclude our study for the case of input thermal states, taking baths at the same temperature, we report in Fig.~\ref{Fig: maximum sLOCC} the maximum success probability $P_\mathrm{LR}^{\max}$, obtained by maximizing Eq.~$($\ref{eq: P_sLOCC}$)$ with respect to $\theta$ under the constraints $F_\mathrm{LR}(\hat{G}^\mathrm{LR}_\mathrm{fin})>2/3$ and $B_\mathrm{LR}(\hat{G}^\mathrm{LR}_\mathrm{fin})>2$, also considering the ideal case of maximum entanglement distillation when $F_\mathrm{LR}(\hat{G}_\mathrm{fin}^\mathrm{LR})=1$ [panel (a)]. We also plot the corresponding optimized spatial deformation parameter $\theta_\mathrm{opt}$ which maximizes the success probability for all the cases [panel (b)]. The fidelity between the output state of Eq.~$($\ref{eq: G thermal}$)$ and the singlet state is in particular given by
\begin{equation}
F_\mathrm{LR}(\hat{G}_\mathrm{fin}^\mathrm{LR})=\frac{2}{3+\cos {(2 \theta)}+4 \cos^2{(\theta)} \cosh (\beta \omega)}.
\end{equation}	
When compared to the ideal case, the curve for the maximum success probability ensuring that the output state violates the Bell inequality shows that the success probability is increased. This result can be further improved considering that, as shown in Ref.~\cite{popescu1994Bell}, Bell inequality violation is a sufficient condition for nonclassical teleportation but is not necessary: in fact, some states fulfilling the necessary condition of $F_\mathrm{LR}(\hat{G}_\mathrm{fin}^\mathrm{LR})>2/3$ do not violate any Bell inequality but are still valuable for teleportation. The curve of $P_\mathrm{LR}^{\max}$ under the condition $F_\mathrm{LR}(\hat{G}_\mathrm{fin}^\mathrm{LR})>2/3$shows that even larger values of the success probability can be now achieved. The curves for the optimized spatial deformation parameters $\theta_\mathrm{opt}$ provide useful information to reproduce experimentally the above results.

\section{Application II: Werner state}\label{SecIV}

We now consider a Werner state \cite{werner1989quantum} as input state, that is the experimentally relevant example of a pure state subject to the action of white noise. In the usual formulation, the Werner state is a mixture of a pure state and the maximally mixed state (white noise). It has been shown that simply bringing identical particles with opposite pseudospins, initially in the elementary state $\ket{\Psi_\mathrm{LR}}=\ket{\mathrm{L}\:\uparrow,\mathrm{R}\:\downarrow}$, to spatially overlap allows to access and generate entanglement by sLOCC measurements \cite{franco2018indistinguishability,sun2020experimental}. However, in a realistic scenario this process might be subject to a certain extent of white noise, modeled by the maximally mixed state $\frac{1}{4}\mathbb{I}_\mathrm{LR}=\frac{1}{4}\sum_{\sigma,\tau=\{\uparrow, \downarrow\}}\ket{\mathrm{L}\:\sigma,\mathrm{R}\:\tau}\bra{\mathrm{L}\:\sigma,\mathrm{R}\:\tau}$ with weight $p$. Under this condition, the input state has the form of the separable (unentangled) Werner state
\begin{equation}
    \hat{W}_\mathrm{int}^\mathrm{LR}=(1-p)\ket{\Psi_\mathrm{LR}}\bra{\Psi_\mathrm{LR}}+\frac{p}{4}\mathbb{I}_\mathrm{LR}.
\end{equation}

\begin{figure}[t!] 
	\centering
	\includegraphics[width=0.48\textwidth]{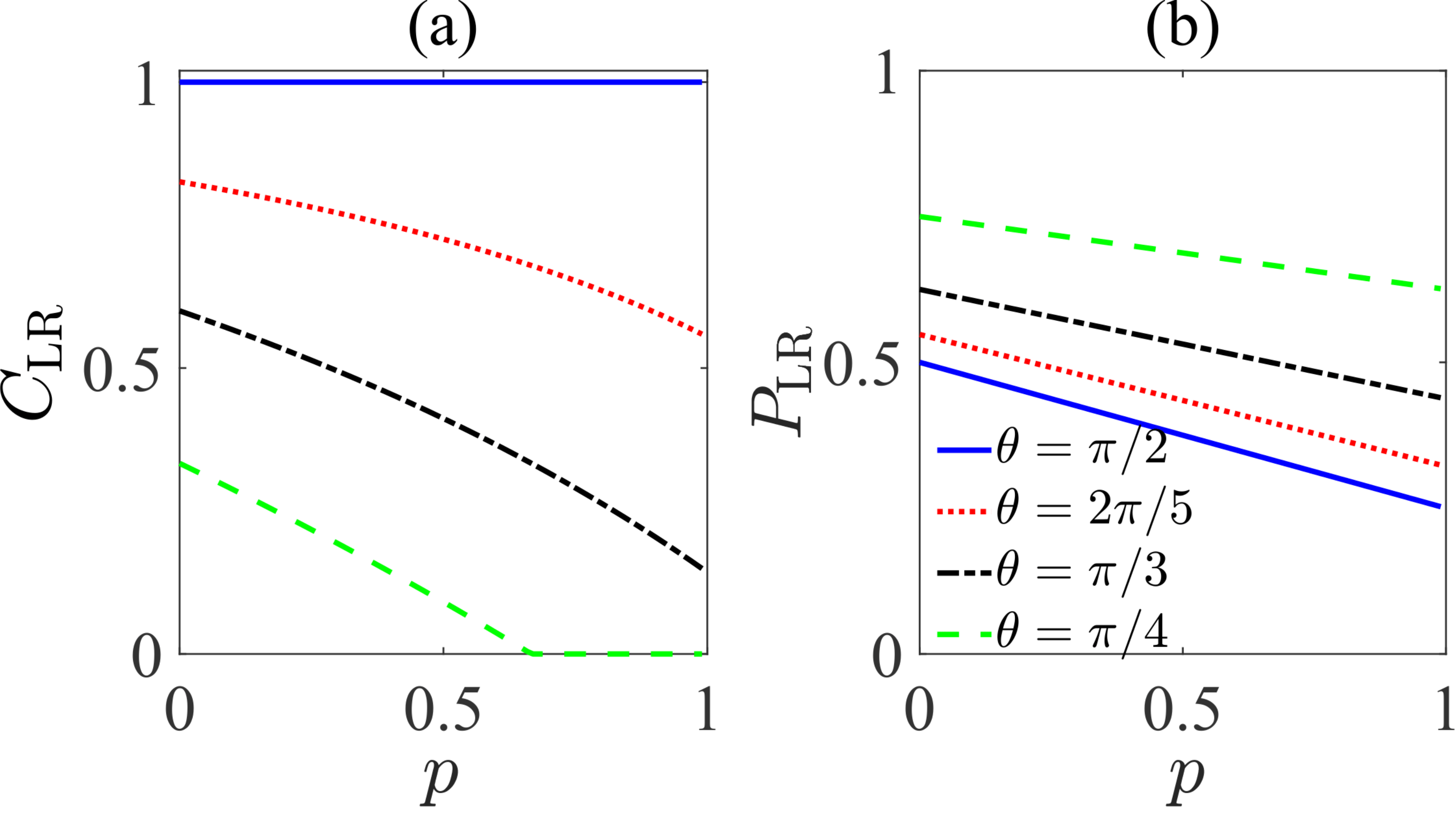}
	\caption{\textbf{Entanglement distillation and success probability  for Werner input states.} (a) Concurrence $C_\mathrm{LR}=C(\hat{W}_\mathrm{fin}^\mathrm{LR})$ and (b) success probability $P_\mathrm{LR}$ as functions of the noise degree $p$ for different values of $\theta$, for bosons.}
	\label{Fig: Thermal entanglement2}
\end{figure}

Following the proposed procedure, we apply the spatial deformation followed by the postselection measurement to distillate the entanglement. The output state is given by $\hat{W}_\mathrm{fin}^\mathrm{LR}=\hat{\Pi}_\mathrm{LR}^{(2)}\hat{W}_\mathrm{D}\hat{\Pi}_\mathrm{LR}^{(2)}/\mathrm{Tr}(\hat{\Pi}_\mathrm{LR}^{(2)}\hat{W}_\mathrm{D})$, with the spatially deformed state $ \hat{W}_\mathrm{D}=\hat{D}(t)\hat{W}_\mathrm{int}^\mathrm{LR}\hat{D}^\dagger(t)$. The corresponding success probability is given by $P_\mathrm{LR}=\frac{1}{8}\left[6-p+(p+2) \cos{(2 \theta)}\right]$. The explicit expression of the output state $\hat{W}_\mathrm{fin}^\mathrm{LR}$ can be straightforwardly obtained by using the general Eqs.~\ref{eq: rho_seprable} and \ref{eq: rho_LR}, and the structure of the coefficients reported in Appendix \ref{appendix2}, where some indications to derive the output state are given. 

For bosons, when $\theta=\pi/2$, the maximum entanglement distillation to the singlet state is possible for any noise degree with a success probability $P_\mathrm{LR}=\frac{2-p}{4}$ ranging from $1/4$ (maximum noise, $p=1$) to $1/2$ (no noise, $p=0$). 
Differently, for fermions, maximum entanglement distillation is only possible when there is no noise ($p=0$), while generally the amount of distilled entanglement is noise dependent. 

For the bosonic case, in Fig.~\ref{Fig: Thermal entanglement2} we plot the degree of distilled entanglement, quantified by the concurrence  $C_\mathrm{LR}$ [panel (a)], and its corresponding success probability $P_\mathrm{LR}$ [panel (b)] as functions of the noise parameter $p$ for different values of $\theta$. We find that to distill a given degree of entanglement from a Werner state with a high amount of noise (e.g., $p>0.5$), a spatial deformation more equally distributed onto the two sites ($\theta$ closer to $\pi/2$) may be required, leading to a larger spatial overlap. This property comes at the cost of a smaller success probability.

\begin{figure}[t!] 
	\centering
	\includegraphics[width=0.48\textwidth]{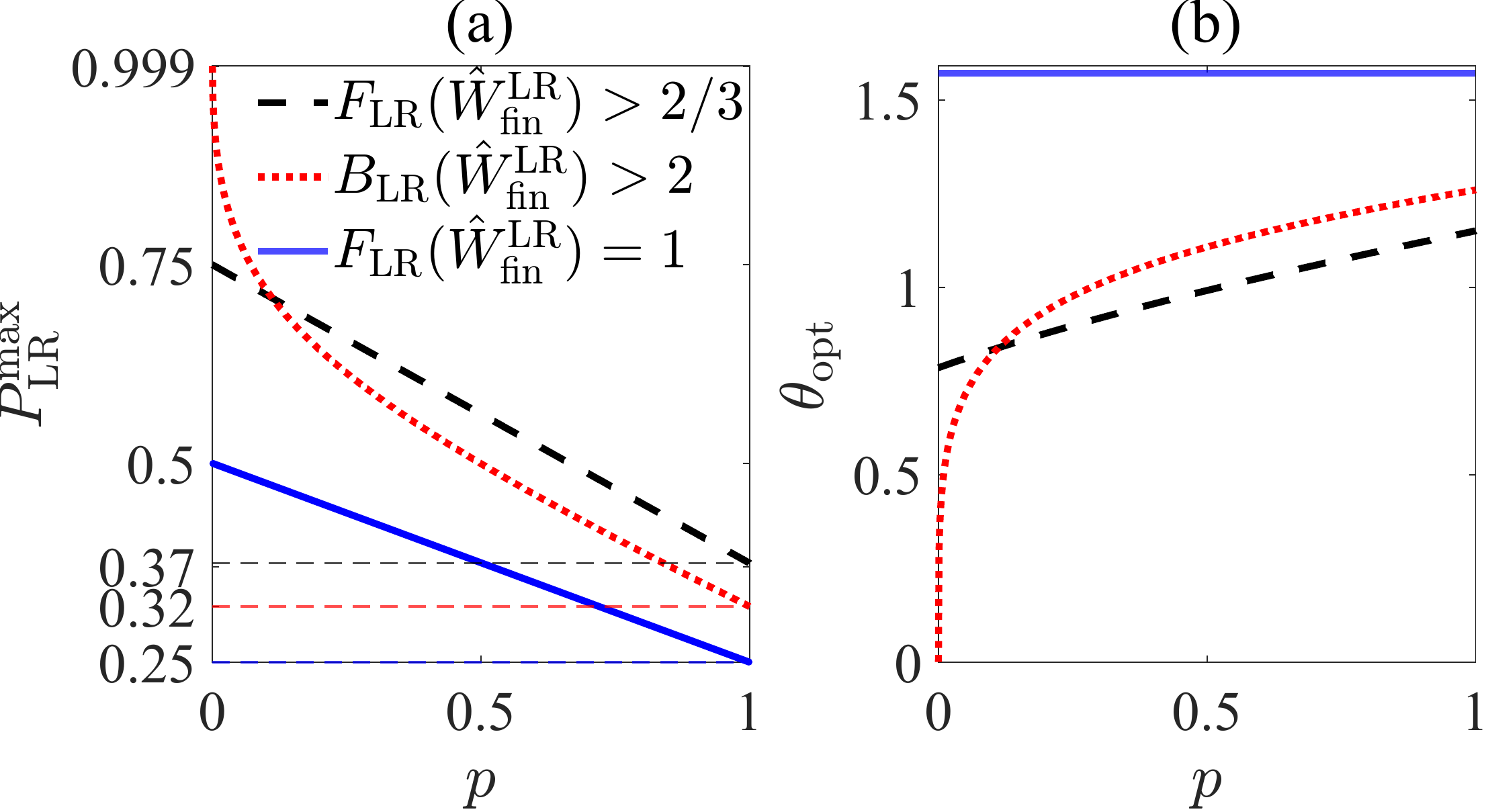}
	\caption{\textbf{Maximum success probability of entanglement distillation with Werner input states.} Maximum  success probability $P_\mathrm{LR}^{\max}$ as a function of the noise degree $p$ for bosons subject to the constraints $F_\mathrm{LR}(\hat{W}^\mathrm{LR}_{\mathrm{fin}})>2/3$ (blue solid line) and $B_\mathrm{LR}(\hat{W}^\mathrm{LR}_{\mathrm{fin}})>2$ (red dotted line). The black dashed line shows the success probability in the case of perfect entanglement distillation, $P_\mathrm{LR}=(2-p)/{4}$. (b) The corresponding optimized spatial deformation parameters $\theta_\mathrm{opt}$ as a function of $p$.}
	\label{Fig: maximum sLOCC Werner}
\end{figure}

We now study the maximization of the success probability with respect to $\theta$ under the desired constraints $ F_\mathrm{LR}(\hat{W}_\mathrm{fin}^\mathrm{LR})>2/3$ and $ B_\mathrm{LR}(\hat{W}_\mathrm{fin}^\mathrm{LR})>2$. In particular, the expression for the fidelity between the output state $\hat{W}_\mathrm{fin}^\mathrm{LR}$ and the singlet state is
\begin{equation}
	F_\mathrm{LR}(\hat{W}_\mathrm{fin}^\mathrm{LR})=\frac{2 (2-p)}{6-p+(p+2) \cos{(2 \theta)}}.
\end{equation}
In Fig.~\ref{Fig: maximum sLOCC Werner}, we plot the maximum success probability $P_\mathrm{LR}^{\max}$ under the two desired conditions for fidelity and Bell inequality violation as a function of the noise parameter $p$, also considering the ideal case of maximum entanglement distillation when $F_\mathrm{LR}(\hat{G}_\mathrm{fin}^\mathrm{LR})=1$ [panel (a)]. We also display the optimized spatial deformation parameter $\theta_\mathrm{opt}$ which maximizes the success probability for all the cases [panel (b)]. We see how the distillation performance depends on the amount of noise in the initial state, pointing out that relatively high probabilities of success can be obtained even for large values of the noise parameter. 
Looking at the curve associated to the violation of the CHSH-Bell inequality, we point out that the maximum success probability $P_\mathrm{LR}^\mathrm{max}$ becomes very high for very low values of the noise degree ($p\rightarrow0$), with the output state resulting close to a pure entangled state for $p\neq0$, and being a pure non-maximally entangled Bell-like state for $p=0$. This fact has a relevant operational consequence within the proposed indistinguishability-assisted entanglement distillation: under weak white noise, small spatial deformations ($\theta\gtrsim 0$), and thus small amounts of spatial overlap between the qubits, are sufficient to efficiently prepare an entangled state violating the Bell inequality (as one can see from the corresponding curve for the optimized spatial deformation parameter $\theta_\mathrm{opt}$). Such an observation is in line with the theorem that any pure Bell-like state  violates a Bell inequality \cite{gisin1991Bell}. From Fig.~\ref{Fig: maximum sLOCC Werner} we also see that $P_\mathrm{LR}^\mathrm{max}$ reaches the minimum value of approximately $0.32$ when the input state is a maximal mixture ($p=1$).  After comparing with to the ideal case ($F_\mathrm{LR}(\hat{W}_\mathrm{fin}^\mathrm{LR})=1$), one can clearly see how much we can increase the success probability, yet guaranteeing that the distilled entangled state is a resource for quantum information processing such as teleportation. The curve of $P_\mathrm{LR}^{\max}$ under the condition $F_\mathrm{LR}(\hat{G}_\mathrm{fin}^\mathrm{LR})>2/3$ shows in particular that a minimum value of approximately $0.37$ can be obtained for $p=1$. Once again, the curves for the optimized spatial deformation parameters $\theta_\mathrm{opt}$ can be useful from an experimental point of view.

\section{Conclusions}\label{SecV}

In this work, we have introduced an entanglement distillation protocol enabled by indistinguishability effects originating from particle statistics and spatial overlap of two identical qubits. The spatial overlap is created by means of a spatial deformation of the wave functions which are distributed towards two distinct measurement sites. The protocol is characterized by a success probability, linked to postselecting the cases where only one particle is found at each site.

We have shown that maximum entanglement distillation is always possible for bosons provided that the initial input mixed state contains at least one term where the two qubits have opposite pseudospins, and determined the corresponding success probability. We have then studied the trade-off between the amount of distilled entanglement and the success probability of the procedure using for the initial configuration thermal Gibbs states and Werner states. We have maximized the success probability with respect to the spatial deformation parameter $\theta$ at the expense of a lower degree of entanglement which yet guarantees Bell inequality violation and teleportation usefulness. The latter is identified using a threshold based on the fidelity of the output state to the maximally entangled singlet state. This analysis is particularly relevant in practical scenarios where efficiency optimization is required.

For the case of initial thermal states we have found that increasing the common temperature of the two local baths is beneficial to the distillation protocol. In the case of different local temperatures, increasing their difference can be also exploited to augment the final entanglement. For the case of initial Werner states, we have found that entanglement distillation can be obtained also for any values of the noise parameter $p$ by increasing the degree of indistinguishability at the expense of a lower success probability.

It is worth mentioning that our protocol is similar to a local filtering protocol \cite{GISIN1996151Hidden} which allows entanglement distillation. However, we remark that the filtering process is here played by the controllable spatial overlap of identical particles and their statistical nature, that is by indistinguishability effects. 

The proposed procedure is amenable to be implemented in the lab, especially in quantum optics platforms. In fact, several experimental tests have been recently realized which exploit the required indistinguishability effects via photons and optical devices \cite{sun2020experimental,barros2020entangling,Sun2022Activation,Wang2022Remote,
lee2022entangling,Wang2022Proof}. Based on the theoretical scheme of Fig.~\ref{Fig_1}, the spatial deformation is a two-mode unitary gate produced by a conventional beamsplitter. Hence, the beamsplitter is why photon bunching, or Hong-Ou-Mandel effect \cite{hong1987measurement}, occurs when both photons are indistinguishable in all their degrees of freedom (polarization, temporal, and so on). Since photons are temporally indistinguishable, they simultaneously arrive at the detectors in the two separated regions, where we only postselect one photon in each region (by means of sLOCC) to generate the entanglement. 

However, as demonstrated experimentally in Ref.~\cite{sun2020experimental}, photons do not always need to cross paths at the same beamsplitter. Depending on the input state, a setup can be assembled where the optical paths of the photons remain detached along their travel until they reach the detection regions where the photons eventually spatially overlap. In such a scenario, two beamsplitters independently distribute each particle wave packet towards two remote operational sites \cite{sun2020experimental,Wang2022Proof} according to the spatial deformations $\hat{D}_{\varphi}(\theta)\ket{0}=\cos{\left(\frac{\theta}{2}\right)}\ket{\varphi_\mathrm{L}}+\mathbf{i}e^{\mathbf{i}\phi}\sin{\left(\frac{\theta}{2}\right)}\ket{\varphi_\mathrm{R}}$ and $\hat{D}_{\chi}(\theta')\ket{0}=\mathbf{i}e^{\mathbf{i}\phi'}\sin{\left(\frac{\theta'}{2}\right)}\ket{\chi_\mathrm{L}}+\cos{\left(\frac{\theta'}{2}\right)}\ket{\chi_\mathrm{R}}$, where $\ket{\varphi(\chi)}_\mathrm{L}$ and $\ket{\varphi(\chi)}_\mathrm{R}$ are orthogonal spatial modes and  $\hat{D}_{\varphi(\chi)}(\theta)$ is the corresponding spatial deformation operator. The particles are then indistinguishable for the final local detectors. The sLOCC is then performed and the entanglement gets distilled. These optical experimental settings can also simulate different particle statistics \cite{tschernig2021direct,Wang2022Proof}, such as fermions and anyons. The above apparatuses are then well suited to design experiments which realize the indistinguishability-assisted entanglement distillation not only for bosonic qubits via photons but also for simulated fermionic particles. Such achievements may open the way to developing quantum repeaters in quantum networks based on indistinguishability effects.    


\section*{Acknowledgments}
R.L.F. acknowledges support from European Union -- NextGenerationEU -- grant MUR D.M. 737/2021 -- research project ``IRISQ''. G.D.C. acknowledge the support by the UK EPSRC EP/S02994X/1 and the Royal Society IEC\textbackslash{R}2\textbackslash{2}22003. F.N. and R.L.F. would like to thank Erika Andersson for useful insights. B.B. thanks Nicol\`{o} Piccione for interesting discussions on the topics of this paper.

\appendix

\section{\label{appendix1} Details on the spatial deformation step}

Let us consider a generic state with a qubit with pseudospin $\sigma$ and another one with pseudospin $\tau$ respectively in regions $\mathrm{L}$ and $\mathrm{R}$: $\ket{\mathrm{L}\:\sigma,\mathrm{R}\:\tau}=\hat{L}_\sigma^{\dagger}\hat{R}_\tau^{\dagger}\ket{0}$, where  $\hat{L}_\sigma^{\dagger}$ ($\hat{R}_\tau^{\dagger}$) is qubit creation operator of with pseudospin $\sigma$ ($\tau$) in region $\mathrm{L}$ ($\mathrm{R}$). In the step I of the procedure, each operator is spatially deformed as $\hat{L}^{\dagger}_\sigma(\theta)=\hat{D}(\theta)\hat{L}_\sigma^{\dagger}\hat{D}^\dagger(\theta)$ and $\hat{R}^{\dagger}_\tau(\theta)=\hat{D}(\theta)\hat{R}_\tau^{\dagger}\hat{D}^\dagger(\theta)$, resulting in 
\begin{eqnarray}
 \hat{L}^{\dagger}_\sigma(\theta)&=&\cos{\left(\frac{\theta}{2}\right)}L_\sigma^{\dagger}+\mathbf{i}e^{\mathbf{i}\phi}\sin{\left(\frac{\theta}{2}\right)}R_\sigma^{\dagger},\nonumber
       \\
       \hat{R}^{\dagger}_\tau(\theta)&=&\mathbf{i}e^{-\mathbf{i}\phi}\sin{\left(\frac{\theta}{2}\right)}L_\tau^{\dagger}+\cos{\left(\frac{\theta}{2}\right)}R_\tau^{\dagger}.
\end{eqnarray}
Therefore, the single-particle states are
$\ket{\psi\:\sigma}=\hat{L}^{\dagger}_\sigma(\theta)\ket{0}$ and $\ket{\psi'\:\tau}=\hat{R}^{\dagger}_\tau(\theta)\ket{0}$. Also, the elementary two-qubit states for different pseudospins ($\sigma\neq\tau$) are, for both fermions and bosons, 
\begin{eqnarray}
    &&\ket{\psi\:\sigma,\psi'\:\tau}=\frac{\mathbf{i}}{2}\sin{(\theta)}\left(e^{-\mathbf{i}\phi}\ket{\mathrm{L}\:\sigma,\mathrm{L}\:\tau}+e^{\mathbf{i}\phi}\ket{\mathrm{R}\:\sigma,\mathrm{R}\:\tau}\right)   \nonumber 
    \\
   && \qquad+  \cos^2{\left(\frac{\theta}{2}\right)}\ket{\mathrm{L}\:\sigma,\mathrm{R}\:\tau}-\eta\sin^2{\left(\frac{\theta}{2}\right)}\ket{\mathrm{L}\:\tau,\mathrm{R}\:\sigma} ,
\end{eqnarray}
which satisfy particle exchange symmetry: $\ket{\mathrm{L}\:\sigma,\mathrm{R}\:\tau}=\eta\ket{\mathrm{R}\:\tau,\mathrm{L}\:\sigma}$. However, when qubits' pseudospins are identical ($\sigma=\tau$), the situation drastically changes depending on particle statistics. For example, when we have fermions the qubits anti-bunch in the state $\hat{L}_\sigma^{\dagger}(\theta)\hat{R}_\sigma^{\dagger}(\theta)\ket{0}=\ket{\mathrm{L}\:\sigma,\mathrm{R}\:\sigma}$, where each qubit resides in a different site. Differently, the elementary two-particle state for bosons becomes
\begin{eqnarray}
        \ket{\psi\:\sigma,\psi'\:\sigma}&=&\frac{\mathbf{i}}{\sqrt{2}}\sin{(\theta)}\left(e^{-\mathbf{i}\phi}\ket{\mathrm{L}\:\sigma,\mathrm{L}\:\sigma}+e^{\mathbf{i}\phi}\ket{\mathrm{R}\:\sigma,\mathrm{R}\:\sigma}\right)\nonumber\\
&&+\cos{(\theta)}\ket{\mathrm{L}\:\sigma,\mathrm{R}\:\sigma},
\end{eqnarray}
where bosonic bunching happens for $\theta=\pi/2$ since both particles are in the same regions. Taking into account these considerations allows for deriving every element of the deformed state in Eq.~$($\ref{eq: deformed_state}$)$. 

\section{\label{appendix2} Details on the activation step}
In the activation step (step II of the procedure), we select only one particle per region using the projective measurement $\hat{\Pi}_{\mathrm{LR}}^{(2)}$. The nonzero probability coefficients in Eq.~$($\ref{eq: rho_LR}$)$ for fermions are
\begin{eqnarray}\label{eq: B1}
\Lambda_{\uparrow,\uparrow}^{\uparrow,\uparrow}&=&\frac{1}{P_{\mathrm{LR}}}\lambda_{\uparrow,\uparrow}^{\mathrm{LR}},\quad 
\Lambda_{\downarrow,\downarrow}^{\downarrow,\downarrow}=\frac{1}{P_{\mathrm{LR}}}\lambda_{\downarrow,\downarrow}^{\mathrm{LR}},\nonumber
\\
\Lambda_{\uparrow,\downarrow}^{\uparrow,\downarrow}&=&\frac{1}{P_{\mathrm{LR}}}\left[\lambda_{\uparrow,\downarrow}^{\mathrm{LR}}\cos^4{\left(\frac{\theta}{2}\right)}+\lambda_{\downarrow,\uparrow}^{\mathrm{LR}}\sin^4{\left(\frac{\theta}{2}\right)}\right],\nonumber
\\
\Lambda_{\downarrow,\uparrow}^{\downarrow,\uparrow}&=&\frac{1}{P_{\mathrm{LR}}}\left[\Lambda_{\downarrow,\uparrow}^{\mathrm{LR}}\cos^4{\left(\frac{\theta}{2}\right)}+\lambda_{\uparrow,\downarrow}^{\mathrm{LR}}\sin^4{\left(\frac{\theta}{2}\right)}\right],\nonumber
\\
\Lambda_{\uparrow,\downarrow}^{\downarrow,\uparrow}&=&\frac{1}{4P_{\mathrm{LR}}}\left[\sin{(\theta)}\left(\lambda_{\uparrow,\downarrow}^{\mathrm{LR}}+\lambda_{\downarrow,\uparrow}^{\mathrm{LR}}\right)\right],
\end{eqnarray}
\\ with $\Lambda_{\uparrow,\downarrow}^{\downarrow,\uparrow}=\Lambda_{\downarrow,\uparrow}^{\uparrow,\downarrow}$ and thesuccess probability is $P_{\mathrm{LR}}=1-\frac{1}{2}\sin{(\theta)}(\lambda_{\uparrow,\downarrow}^{\mathrm{LR}}+\lambda_{\downarrow,\uparrow}^{\mathrm{LR}})$. Also, the nonzero bosonic probability coefficients of Eq.~$($\ref{eq: rho_LR}$)$ are 
\begin{eqnarray}\label{eq: B2}
\Lambda_{\uparrow,\uparrow}^{\uparrow,\uparrow}&=&\frac{1}{P_{\mathrm{LR}}}\lambda_{\uparrow,\uparrow}^{\mathrm{LR}}\cos{(\theta)},\quad
\Lambda_{\downarrow,\downarrow}^{\downarrow,\downarrow}=\frac{1}{P_{\mathrm{LR}}}\lambda_{\downarrow,\downarrow}^{\mathrm{LR}}\cos{(\theta)},\nonumber
\\
\Lambda_{\uparrow,\downarrow}^{\uparrow,\downarrow}&=&\frac{1}{P_{\mathrm{LR}}}\left[\lambda_{\uparrow,\downarrow}^{\mathrm{LR}}\cos^4{\left(\frac{\theta}{2}\right)}+\lambda_{\downarrow,\uparrow}^{\mathrm{LR}}\sin^4{\left(\frac{\theta}{2}\right)}\right],\nonumber
\\
\Lambda_{\downarrow,\uparrow}^{\downarrow,\uparrow}&=&\frac{1}{P_{\mathrm{LR}}}\left[\lambda_{\downarrow,\uparrow}^{\mathrm{LR}}\cos^4{\left(\frac{\theta}{2}\right)}+\lambda_{\uparrow,\downarrow}^{\mathrm{LR}}\sin^4{\left(\frac{\theta}{2}\right)}\right],\nonumber
\\
\Lambda_{\uparrow,\downarrow}^{\downarrow,\uparrow}&=&\frac{-1}{4P_{\mathrm{LR}}}\left[\sin{(\theta)}\left(\lambda_{\uparrow,\downarrow}^{\mathrm{LR}}+\lambda_{\downarrow,\uparrow}^{\mathrm{LR}}\right)\right],
\end{eqnarray}
with $\Lambda_{\uparrow,\downarrow}^{\downarrow,\uparrow}=\Lambda_{\downarrow,\uparrow}^{\uparrow,\downarrow}$ and thesuccess probability is $P_{\mathrm{LR}}=1-\frac{1}{2}\sin{(\theta)}\left(1+\lambda_{\uparrow,\uparrow}^{\mathrm{LR}}+\lambda_{\downarrow,\downarrow}^{\mathrm{LR}}\right)$. 

It is worth mentioning that the derivation of Eq.~$($\ref{eq: G thermal}$)$ involves substituting the following thermal coefficients in Eq.~$($\ref{eq: B1}$)$ and Eq.~$($\ref{eq: B2}$)$: $\lambda_{\uparrow,\uparrow}^{\mathrm{LR}}=e^{\beta\omega}$,  $\lambda_{\uparrow,\downarrow}^{\mathrm{LR}}=1$, and $\lambda_{\downarrow,\uparrow}^{\mathrm{LR}}=1$, and $\lambda_{\downarrow,\downarrow}^{\mathrm{LR}}=e^{-\beta\omega}$. After diagonalization, Eq.~$($\ref{eq: G thermal}$)$ can be obtained. This approach is also applicable to the Werner state, where the substitutions $\lambda_{\uparrow,\uparrow}^{\mathrm{LR}}=p/4$, $\lambda_{\uparrow,\downarrow}^{\mathrm{LR}}=1-3p/4$, and $\lambda_{\downarrow,\uparrow}^{\mathrm{LR}}=p/4$, and $\lambda_{\downarrow,\downarrow}^{\mathrm{LR}}=p/4$ lead to the final state for entanglement distillation.


%

\end{document}